# Legacy Forensics: An Emerging Challenge

Thomas P. Dover[1]

**ABSTRACT**

With the passage of time and as new types of storage devices are introduced into the marketplace, contemporary devices will slowly lose their compatibility with current operating systems and PC hardware. As a result, such "legacy" devices will pose an analytical challenge to the field of digital forensics. Dated technology, while still fully functional, is becoming increasingly incompatible with most contemporary computing hardware and software and thus cannot be properly examined in today's digital forensic environment. This fact will not be lost on those who utilize legacy hardware (and software) when committing criminal acts. This paper describes the technical challenge of accessing legacy devices by describing an effort to resuscitate a Bernoulli Drive; a portable storage device manufactured in 1983 by Iomega Corporation. A number of "lessons learned" are provided and the implication of legacy devices to digital forensic science is discussed.

---

[1] Adjunct Faculty, Business & Technologies, Community College of Beaver County, Monaca, PA, thomas.dover@ccbc.edu

**INTRODUCTION**

There it sat, my old Bernoulli Drive[2], (figure 1) the "portable" storage device [1] I had used in the early 1990's to store digital evidence[3]. For over 20 years it had sat in a basement storage box untouched with no cables, no connectors and no manual to be found. I couldn't even remember what type of computer it had originally been connected to. It did, however, still have a 230mb floppy "diskette" inserted into its drive bay.

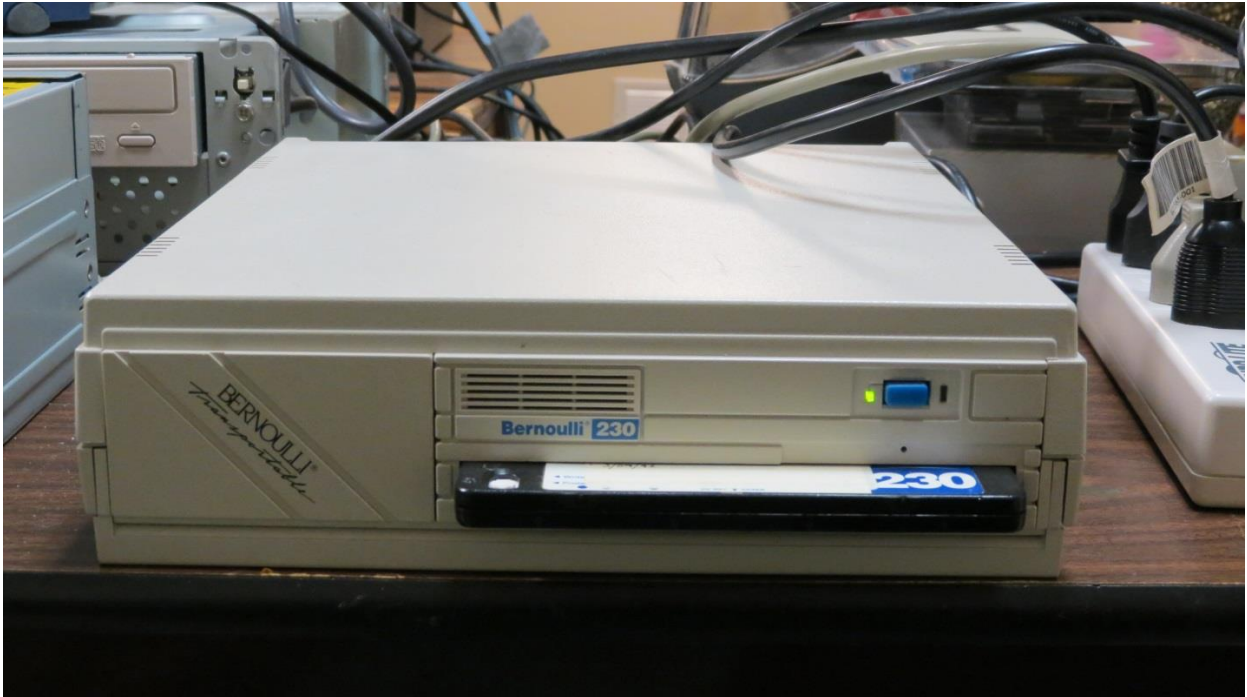

Figure 1 (Bernoulli Drive model B1230T)

Staring at the drive made me wonder. What if, as a digital forensic examiner, I was presented with such an *old* device for examination? Unless physically damaged it should (be expected to) function as designed. After all, I thought, it (like the floppy diskette) was not made obsolete due to any flaw in design or engineering. If this was the case then there was no reason, provided I could get it to work, why the Bernoulli Drive could not continue to read and write data just like any other storage medium.

By virtue of vintage, obsolete technology poses a unique challenge to contemporary digital forensics. Given the ubiquity of flash drives, with the exception of external storage[4] almost all other forms of portable data storage have been relegated to obsolescence or history[5].

Digital (or computer) forensics, as a scientific discipline, is little more than twenty years old and not much has been written about how the passage of time impacts (or will impact) the ability of examiners to keep pace with new forms of data storage while retaining the historical knowledge, skills and equipment needed to analyze "legacy" devices in a forensically sound manner. Indeed, current standards and accepted practices for acquisition and analysis may not even be applicable to such devices.

---

[2] The Bernoulli drive (named after Daniel Bernoulli) was manufactured by Iomega Corporation and first produced in 1982. The model shown in figure 1 is B1230T.
[3] The Bernoulli drive supplemented my Axonix Corporation (Salt Lake City, UT) 200mb portable hard drive for digital evidence storage (circa 1989-1992).
[4] i.e., external, portable hard drives.
[5] e.g., floppy diskettes, Zip disks, CDs, PDAs

Defining what qualifies as a "legacy" device is difficult given the axiom "if it's still being used it can't be a legacy." For the purpose of this paper, however, hardware or software is defined as "legacy" if it meets any of the following criteria:

1. It is, on average, more than 10 years old
2. It is two or more generations removed from its successor
3. For hardware, it is unable to run contemporary software
4. For software, it cannot be run on contemporary hardware
5. It is no longer supported by its original designer or manufacturer

The ability of today's digital forensic examiners to successfully test old technology is inhibited by the general observation that most of today's PC's used for digital forensics are simply commercially available desktops or towers with expanded hardware[6] and dedicated forensic software. Disappearing from modern computers are serial and parallel ports, RGB video connector and floppy diskette drives[7]. In some computers even CD/DVD-ROM drives are considered add-ons. Even high-performance, dedicated forensic workstations, such as Digital Intelligence's FRED [2] platform includes neither parallel nor serial ports.[8] For general computing this type of progress is beneficial, however, from a digital forensic perspective it negatively impacts forensic analysis of an entire class of (still functional) data storage devices. This impact will not be lost on those who will use whatever technology is available (old or new) in furtherance of criminal activity.

While it is not impossible to read and forensically image data from obsolete technology such as floppy diskettes, Bernoulli Drives and zip disks[9], it does require a computing platform with the requisite hardware and software for such effort to be successful. Fortunately, some forensic software remains capable of imaging legacy devices such as floppy diskettes or any other logical drive connected to a forensic platform. For example, Access Data's *Forensic Toolkit Imager* program will create logical images of floppy diskettes in a contemporary OS environment.

It can be argued that the issue of legacy device forensics is neither significant nor important enough to warrant serious consideration by the digital forensic community. One possible reason for this belief is that there are insufficient instances of this type of examination being reported to justify the expenditure of time, staff and funding needed to create a legacy environment suitable for forensic analysis.

According to a survey conducted by Computer Forensic World.com [3] less than 10% of those polled[10] claimed "more than 5 years of experience in the forensics industry." If this number is valid it also means that most examiners have never dealt with legacy devices. Moreover, it is uncertain if digital forensic labs are outfitted with the type of hardware and software needed to support such examinations. Of course, an argument can be made that the simplest and most direct solution to a legacy device is to simply run it on whatever computer it was connected to[11]. While doing so violates a long-held digital forensic principle[12], it also assumes that the computing environment it was operating in when seized can be duplicated.

Back to the Bernoulli Drive, however. After giving the matter further thought, I decided to see if I could, in fact, resuscitate a two decade old piece of storage technology well enough to view its contents. If successful, it would validate my suspicion that given the correct platform and proper configuration legacy devices remain just as capable of storing data now as when they were first introduced.

---

[6] Maximum RAM and video memory, multiple hard drives and extra expansion slots.
[7] Replaced by USB, Firewire, DVI and/or HDMI ports.
[8] Per FRED configuration specifications.
[9] Iomega has updated older zip drives (such as the parallel port zip drive) to support USB technology. The company's newer Zip drives are capable of reading older 100 & 250mb zip disks.
[10] Approximately 16,000 respondents
[11] The assumption is that the legacy device was seized relative to a consent search or search warrant.
[12] Do not examine digital evidence using seized equipment.

## THE CHALLENGE (OR "HOW I DID IT"[13])

The first step was to determine if the Bernoulli would even power up. Fortunately, its power connector is 110v AC so any standard 3-prong power cord would work. Not surprisingly, the Bernoulli powered up as if it were 1990; the whirring of the drive head audible as it engaged the still inserted floppy disk. A few seconds later the activity lights indicated the device was operational and in ready-mode. The Eject button worked as it always had and the 230 megabyte diskette released itself from the drive bay just like it had done two decades ago. I was now in possession of a *time machine*. If it could be connected to a correctly configured PC I would, in effect, be able to go back "in time" and see what files had been stored on that diskette[14].

With the Bernoulli successfully powered up the next step was determining what kind of cable connections it needed along with what PC configuration would support it. First stop, Iomega Corporation's web site[15]. Sadly, the Bernoulli home page consisted of a single page indicating that service and support for Bernoulli drives was now handled by another vendor (Comet Enterprises). The page listed a telephone number and web URL for the company.

Before attempting to contact Comet Enterprises, a fair number of web searches and Wikipedia entries provided me with bits and pieces about the Drive itself. I (re)learned that my Bernoulli was a SCSI drive which meant I needed a SCSI adapter card [4][16] (figure 2). I also needed a SCSI-1 cable[17] for the device-to-PC connection. Obtaining these two pieces of hardware completed the physical-connection portion of the challenge. I also decided to obtain a SCSI terminator although I wasn't sure if it would be needed or required by the Bernoulli for proper functioning.

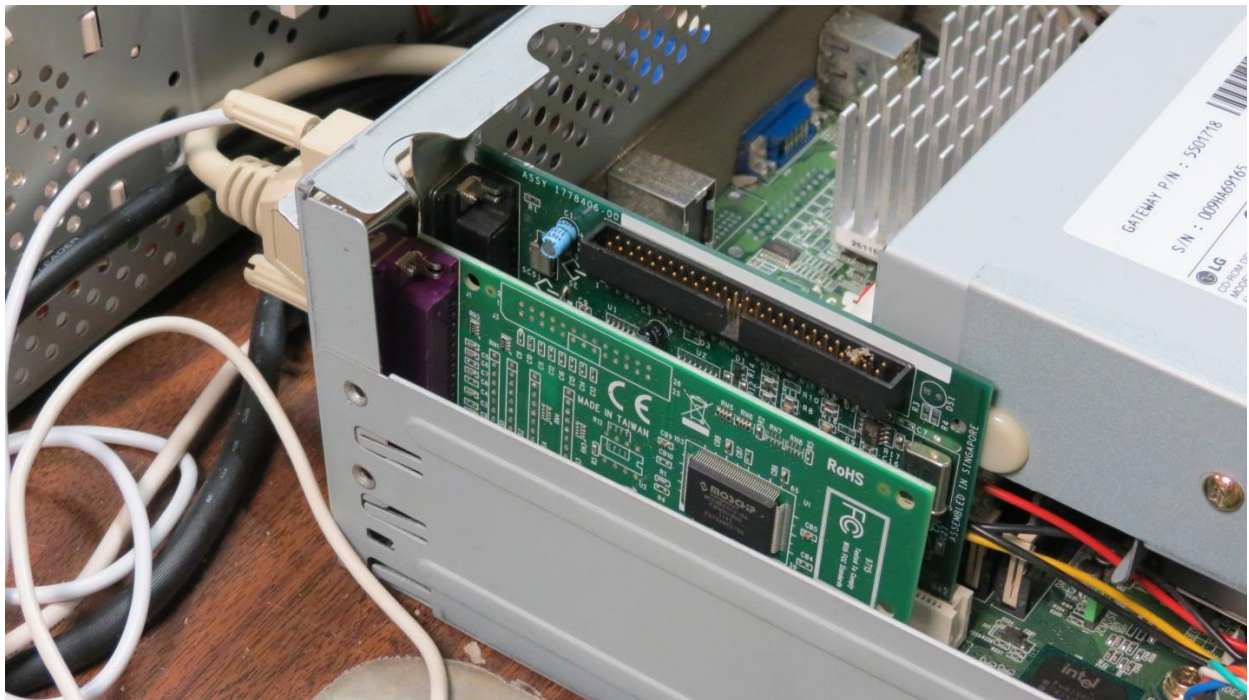

Figure 2 (Adaptec PCI SCSI Card / rear)

---

[13] Apologies to the 1974 Mel Brooks movie "Young Frankenstein"
[14] A sentiment probably shared by anyone having such old storage mediums lying around without the means to read them.
[15] http://www.iomega.com
[16] Adaptec SCSI card, model 2906, which included an external db25 pin connector. Importantly, this model supported DOS versions 3-6 and Windows through XP.
[17] Centronics DB50 M/CN50 M SCSI cable.

Next, I had to select a suitable PC platform and operating system. Ideally, it should be a PC from the Bernoulli's generation[18] or close to it. I opted, initially, to use my test platform; a 10 year-old Gateway mini-tower[19] running Windows7-64. While the Gateway's standard PCI slot would accommodate the SCSI card without issue the card I needed and eventually chose was not supported by any version of Windows beyond XP. This conflict ruled out my test platform.

I turned once again to the storage boxes in my basement and identified a couple of desktops that were possibilities. The first candidate was a Compaq desktop, circa 1995 and the second a Gateway model 866, circa 2000 (figure 3). Neither computer had been powered up for at least 10 years. After pulling both from storage, cleaning off the dust and making a couple of safety checks each was cautiously connected to power. As with the Bernoulli, both the Compaq and Gateway powered on just as they had years ago. The Gateway 866 was selected for the Bernoulli as its PCI card slot was easier to reach than the Compaq's. It also had more memory and a larger hard drive which would enhance the chances of success. The Compaq was still running Windows 3.1 which was left installed and due to its parallel port (which the Gateway 866 did not have) another legacy device, an Iomega Zip Drive (parallel port model)[20] was connected and with the appropriate drivers (which can still be downloaded from Iomega) loaded it was successfully recognized by the operating system.

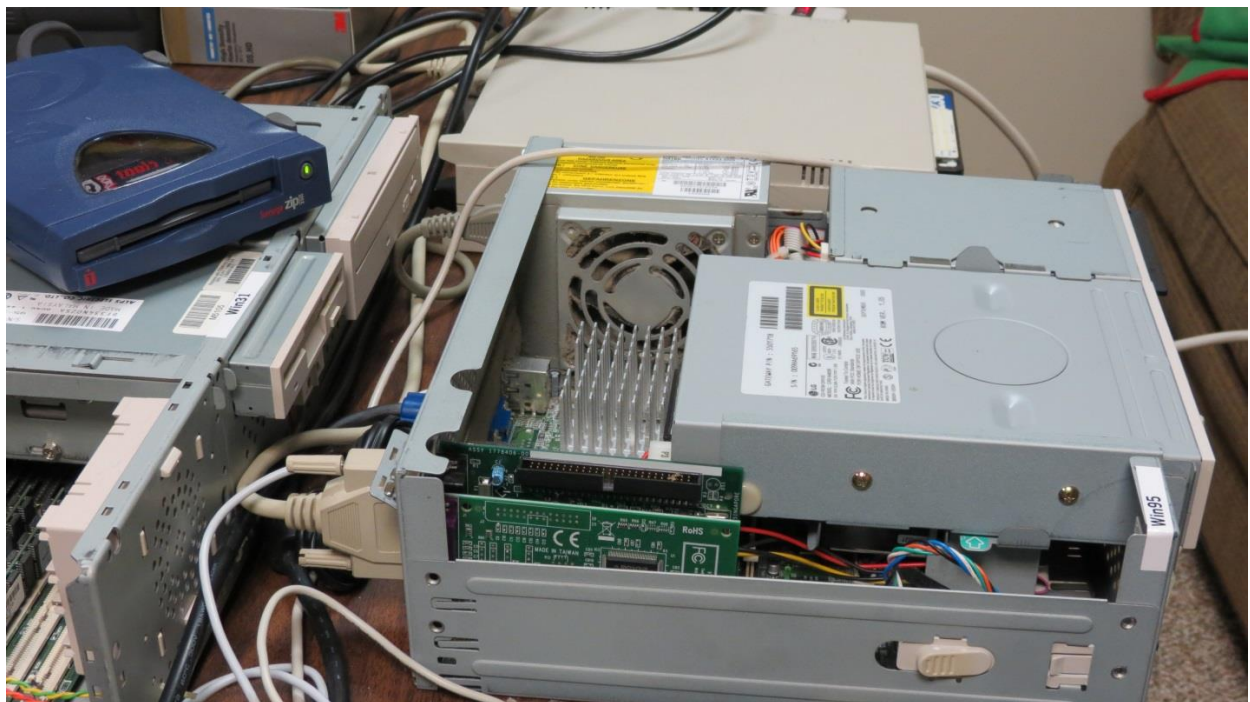

Figure 3 (Gateway PC model 866)

---

[18] Or at least 5-7 years on either side of its manufacture date.
[19] It had been given to me by an associate who was upgrading. After maxing out its memory, installing a new video card and upgrading the hard drive I had an optimal forensic test platform.
[20] Iomega produced this version of Zip Drive in the early to mid-1990s and it can be seen in the upper left corner of Figure 3.

With the Adaptec card installed I turned my attention to the Gateway's operating system. It was originally purchased with Windows 2000 pre-installed. Unfortunately, the Bernoulli Drive was not supported after Windows 95 and it wasn't until Windows 95 (versions B & C) that the USB 1.0 specification [5] was supported. Once again, I turned to those basement storage boxes. A repository of operating systems, applications and utilities (in 3.5" & 5.25" floppy disk and CD format) including all versions of DOS from 3.0-6.0 and Windows from 3.1 thru Win7 did the trick. After formatting and re-partitioning the hard drive I installed DOS 6.0 followed by Windows 3.1 and finally Windows 95. Installation went without a hitch and I now had a fully functioning Windows 95 desktop computer. Next, the drivers for the Adaptec SCSI card were loaded[21] and the last step (which would be the most challenging) was to locate the drivers for the Bernoulli Drive itself.

As mentioned earlier, Iomega had sold the Bernoulli line[22] to Comet Enterprises. I contacted the company directly but without success. Email was next and in somewhat of a detailed message I asked if the company had Windows 3.1 or 95 drivers for the Bernoulli. Although an initial reply was received (asking for the version of Windows needed) I never heard back from the company despite attempts to follow-up.

Stymied without the correct driver I decided to give Iomega another try. Surely someone, I thought, at the company would either have or know the whereabouts of driver software for one of its very first products. After several emails explaining my plight, "Richard" from Iomega Technical Support provided me with a driver package for Windows 3.1 and Windows 95. Of course, Richard emphasized that it was OLD software no longer supported by the company. Nevertheless, I was grateful for having what I hoped was the last piece of the puzzle needed to make the Bernoulli Drive once again operational.

---

[21] This task required an interim step. The driver software came on CD and the Gateway's CD was not recognized by Windows 95 which meant first copying the driver files from CD to 3.5" floppy diskette. An additional problem presented itself when the files on the CD were larger than the floppy diskette's capacity. This problem was resolved by splitting the files over several diskettes.
[22] Per Iomega Tech Support

After carefully installing the Bernoulli's Windows 95 driver and after three months of intermittent effort (and ~ $100 in hardware) I achieved what I had up until now doubted; an almost 25 year-old storage device, connected to a 15 year-old PC with a 20 year-old operating system was recognized and presented itself as a drive letter in Windows Explorer (Figure 4).

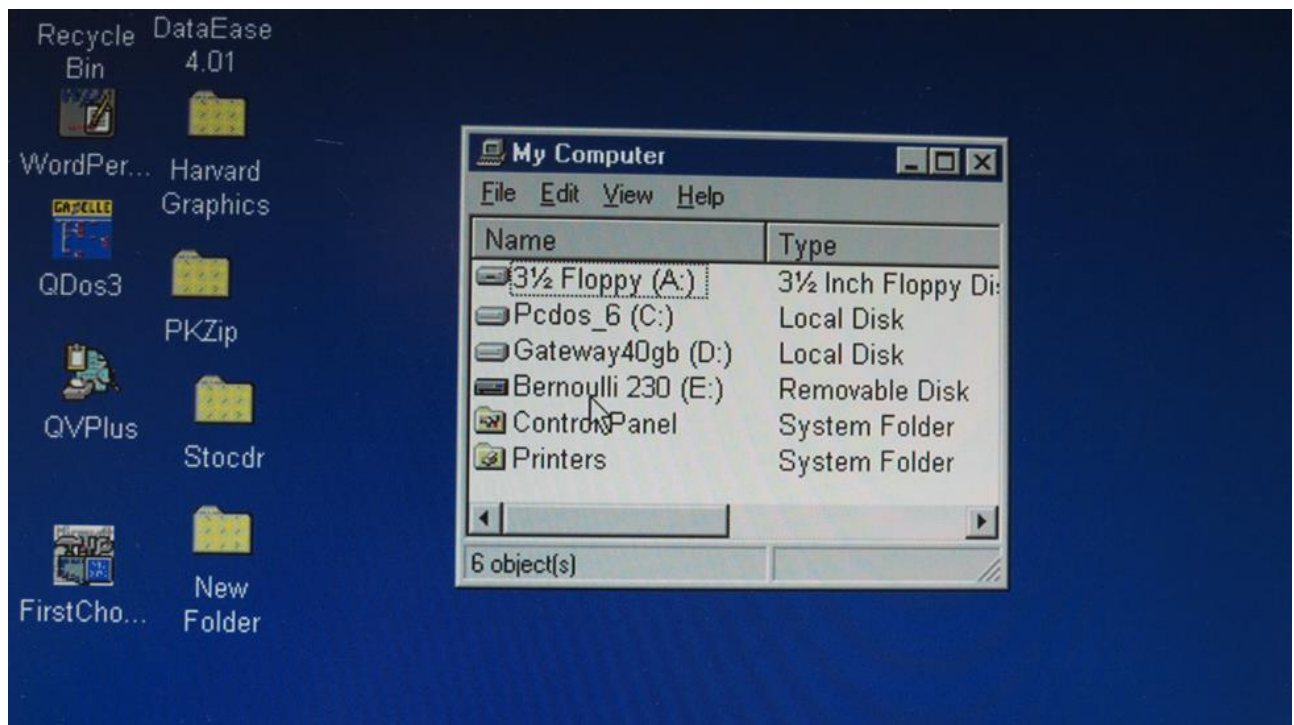

Figure 4 (Bernoulli Drive (E:) in Windows (95) Explorer)

**LESSONS LEARNED**

When assessing digital forensics involving legacy devices the following points should be considered:

1. **Do Not Throw *Everything* Out**. It is common practice for both digital forensic examiners and their supporting labs to replace PCs, cables, connectors, other devices and software as needed. Consider, however, retaining at least one or two working systems which can be mothballed and put into storage until needed. The same consideration goes for software. An ideal configuration would be one or more legacy PCs capable of running dedicated operating systems (e.g., DOS, Win31, Win95, Win2000, etc…).

2. **Keep Copies of Everything**. Every operating system, application, utility or other piece of software should be kept and archived on at least two different types of media (e.g., CD and Flash Drive). This is especially true for legacy applications stored on obsolete storage mediums such as floppy diskettes. This point also applies to serial numbers or registration codes (for applications which require them).

3. **Keep All Driver Versions**. Keep all versions for any hardware device that requires them. This can be a particularly difficult challenge for a device that is no longer supported by the manufacturer or vendor.

4. **Establish a Common Exchange Medium between Legacy and Contemporary PCs**. Once a legacy device can be read it is usually necessary to transfer its contents to another PC for analysis. This action requires establishing a level of exchange that both old and new PC recognize and can share. The most common shared mediums include floppy diskette, Zip disk or CD-ROM. Strive to establish the highest (in terms of volume) level of common exchange possible.

5. **Leave Legacy PCs as Standalone Platforms**. Unless absolutely necessary, there is no reason to connect a legacy PC to the internet. At no time during my effort to resuscitate the Bernoulli Drive was a direct internet connection needed. If internet access was required I used a different PC and transferred the files to the legacy PC via floppy diskette. Most legacy computers will not possess either a NIC card or built-in RJ-45 port[23].

## CONCLUSION AND FUTURE WORK

Legacy devices present a unique set of challenges to the digital forensic community that includes backwards compatibility with contemporary systems, examiner experience working with legacy devices and availability of compatible hardware and software.

Although contemporary PC's and operating systems remain capable of supporting legacy hardware if supported by the USB standard[24], the same cannot be said for legacy devices such as a Bernoulli Drive, (5.25") floppy diskettes[25] or Personal Digital Assistants[26] (PDA).

My effort to resuscitate a 20+ year old piece of obsolete but still functional technology owed its success, in part, to personal ownership of such dated technology. While this paper outlined the steps needed to make a legacy device operational it requires a commitment of time and effort, and a certain amount of funding above what is probably allocated to most routine digital forensic examinations, to be successful. As stated earlier, most forensic labs probably do not possess "legacy" PCs[27] or other hardware/software capable of reading such devices and finding legacy computer equipment (in good working order) is becoming more and more difficult[28].

A concern about technology from a digital forensic perspective should be that today's technology is tomorrow's obsolescence. As time passes and as new storage mediums replace older but still functional ones the spectrum of legacy devices suitable for forensic analysis expands. Unfortunately, obsolete storage mediums cannot be permanently "retired" from use. Today's flash drives may be considered "quaint" storage mediums in ten years but will remain as functional then as now. Like an upside-down pyramid, the various types of devices that can hold data will continue to grow and expand as newer storage technologies are introduced into the marketplace.

As mentioned earlier, this fact will not be lost on the criminal element. While it is unknown to what extent legacy equipment plays a role in today's cyber world, it can be conjectured that if the use of such equipment furthers the aims of the cyber (or non-cyber) criminal then if it can be used, it will be used. Moreover, using legacy devices could even be considered "counter-forensic" since forensic examinations involving legacy devices would require a commitment of resources that digital forensic labs may be unable to support or address.

While the growing cyber-crimes of network breaches, attacks and disruptions (with their attendant media attention) has drawn the focus of many in the digital forensic community, non-volatile or "dead box" examinations remain one of the field's core functions. It is important that those practicing this science maintain their awareness of not only what is emerging but what has been. Of course, the whole issue of legacy devices would be an academic one if they simply were no longer capable of doing what they were designed and manufactured to do. As this paper demonstrates, however, legacy devices are as capable of storing data now as when they were first introduced.

---

[23] Most legacy systems will only have RJ-11 ports.
[24] Examples include (external) 3.5" floppy diskette and Zip drives.
[25] 5.25" floppy diskettes require an IDE controller. External or USB-supported drives are not manufactured.
[26] Examples include HP's iPaq, the Zire and Sharp's Wizard
[27] The general tendency is to replace dated equipment every few years. Such equipment is normally destroyed or otherwise disposed.
[28] eBay is the obvious web site for legacy equipment but Goodwill Industries also has a large selection and there are a number of "vintage" computer equipment sites on the internet.

Finally, this paper did not address the issue of analyzing data retrieved from legacy devices nor did it examine the possibility of using virtual machines as a means to run legacy operating systems well enough to support obsolete hardware and software. Both are questions left for another day.

**REFERENCES**

[1] Iomega Corp., "25 Years," 2013. [Online]. Available: http://www.iomega.com.

[2] Digital Intelligence , "FRED," Digital Intelligence, 2013. [Online]. Available: http://www.digitalintelligence.com/products/fred/. [Accessed 2 March 2013].

[3] Computer Forensics World, "Poll Results," 2013. [Online]. Available: http://www.computerforensicsworld.com/modules.php?name=Surveys&op=results&pollID=1&mode=&order=&thold=. [Accessed 1 April 2013].

[4] Adaptec, "Adaptec SCSI Card 2906," Adaptec, 1996-2013. [Online]. Available: http://www.adaptec.com/en-us/support/scsi/2900/ava-2906/. [Accessed 15 March 2013].

[5] Support, "Availability of Universal Serial Bus Support in Windows 95," Microsoft, 15 November 2006. [Online]. Available: http://support.microsoft.com/kb/253756. [Accessed 12 March 2013].